\pdfoutput=1

\documentclass[aps,prl,twocolumn,superscriptaddress]{revtex4-2}

\usepackage{graphicx}
\usepackage{amsmath, amssymb}

\usepackage{siunitx}
\usepackage{hyperref}

\begin{document}

%Title of paper
\title{\textbf{First Demonstration of Resonant Pitch-Angle Scattering of Relativistic Electrons by Externally-Launched Helicon Waves} } 

\author{H. Choudhury}
\email{Contact author: hpc2111@columbia.edu}
\affiliation{Columbia University, New York, New York 10027, USA}
\author{A. Battey}
\affiliation{EPFL, Swiss Plasma Center (SPC), Lausanne CH-1015, Switzerland}
\author{C. Paz-Soldan}
\affiliation{Columbia University, New York, New York 10027, USA}
\author{J. Lestz}
\affiliation{General Atomics, San Diego, California 92186, USA}
\author{N. Leuthold}
\affiliation{Columbia University, New York, New York 10027, USA}

\author{A. Lvovskiy}
\affiliation{General Atomics, San Diego, California 92186, USA}
\author{C. Marini}
\affiliation{University of California San Diego, La Jolla, California 92093, USA}
\author{J. Barr}
\affiliation{General Atomics, San Diego, California 92186, USA}
\author{W. Heidbrink}
\affiliation{University of California, Irvine, California 92697, USA}

\author{D. Spong}
\affiliation{Oak Ridge National Laboratory, Oak Ridge, Tennessee 37831, USA}
\author{S. Tang}
\affiliation{General Atomics, San Diego, California 92186, USA}
\author{B. Van Compernolle}
\affiliation{General Atomics, San Diego, California 92186, USA}

% \author{Y. Ghai}
% \affiliation{Oak Ridge National Laboratory}
% \author{X. Tang}
% \affiliation{Los Alamos National Laboratory (LANL)}

% \author{G. Bodner}
% \affiliation{General Atomics}
% \author{D. Liu} ask BIlll about this
% \affiliation{General Atomics}

\author{Q. Zhang}
\affiliation{Los Alamos National Laboratory, Los Alamos, New Mexico 87545, USA}
\author{Y. Zhang}
\affiliation{Los Alamos National Laboratory, Los Alamos, New Mexico 87545, USA}
\author{X. Tang}
\affiliation{Los Alamos National Laboratory, Los Alamos, New Mexico 87545, USA}
% \author{S. Ding}
% \affiliation{General Atomics}

%todo \thanks{}

% ask Jeremy Hanson and Jason Barr whether they wanted to be on 

\date{\today}

\begin{abstract}
Helicon waves satisfying the normal wave-particle cyclotron resonance are observed to limit the growth and maximum energy of relativistic electrons (REs) in low-density Ohmic DIII-D tokamak plasmas. Following the application of helicon waves, pitch-angle scattering of high-energy REs causes an increase in both synchrotron and electron-cyclotron emissions. The hard x-ray emission, a proxy for the RE population, ceases to grow; and energy-resolved hard x-ray measurements also show a striking decrease in the number of high-energy REs (above the resonance at approximately \SI{8}{MeV}) to below the noise floor. This occurs despite the toroidal electric field remaining high enough to drive exponential RE growth in the absence of helicon waves. These results open new directions for limiting the maximum energy of RE populations in laboratory and fusion plasmas.
\end{abstract}

\maketitle

\paragraph{Introduction}

Relativistic electrons (REs) pose a formidable risk to the operation of future high-current tokamaks \cite{boozer_runaway_2017, boozer_pivotal_2018}. During a fast, unintended disruption in which REs carry most of the plasma current, an uncontrolled release of REs from the plasma to the wall would cause severe damage to plasma-facing components \cite{boozer_theory_2015,hollmann_status_2014,lehnen_disruptions_2015}.
\par
For a fully-ionized quasi-neutral plasma, relativistic theory predicts that electrons experience an acceleration that exceeds the drag on them whenever the toroidal electric field, $E = V/2\pi R$, in a tokamak of loop voltage $V$ and major radius $R$, exceeds a critical value \cite{connor_relativistic_1975}, 
\begin{equation}
E_C=\frac{n_e e^3 \ln \Lambda}{4 \pi \varepsilon_0^2 m_e c^2},
\label{eq:Ec}
\end{equation}
in a plasma of density $n_e$ and Coulomb logarithm ln$\Lambda$ \cite{freidberg_plasma_2007}. The drag force, mostly from electron-electron coulomb collisions, \textit{decreases} as the electrons are accelerated, prompting them to runaway to relativistic energies \cite{breizman_physics_2019}. In experiments, however, it has been observed that some multiple of $E_C$, typically $\geq5$, must be exceeded for significant RE growth, suggesting energy loss mechanisms other than collisional damping play an important role \cite{paz-soldan_growth_2014, granetz_itpa_2014}. Recent theoretical works have suggested that synchrotron damping and kinetic instabilities are at least partially responsible for this enhancement of  the critical field \cite{aleynikov_theory_2015,liu_role_2018}.
\par
Once there is a seed population of REs, individual REs can collide with thermal electrons, knocking them into the RE regime whilst remaining relativistic themselves - the so-called avalanche process  \cite{rosenbluth_theory_1997}. This process is of particular concern for high-current tokamaks currently under construction like ITER and SPARC since it scales exponentially with the plasma current \cite{boozer_pivotal_2018, greenwald_status_2020}. 
\par
Owing to the risks REs pose, such as damaging plasma-facing components, it is crucial that mitigation strategies be developed to limit the damage an unintended disruption may do. The current strategy for RE-mitigation on ITER involves a massive injection of high-Z impurities (e.g argon or neon) to radiate the magnetic and thermal energy of the REs and to increase the critical field by increasing the density \cite{whyte_mitigation_2002, hollmann_control_2013,boozer_runaway_2017,boozer_theory_2015}. Other mitigation strategies involve MHD effects, such as the use of a passive coil external to the plasma to break up magnetic surfaces during a disruption \cite{boozer_two_2011}. Another involves injecting hydrogen isotopes to excite large-scale current-driven MHD instabilities, which have been shown to benignly terminate RE beams on DIII-D and JET \cite{reux_demonstration_2021, paz-soldan_novel_2021}. In general, mitigation strategies either increase the RE loss rate or increase $E_C$ to suppress their generation. 
\par
A novel approach to RE mitigation involves the use of externally launched radio-frequency waves to pitch-angle scatter REs through resonant wave-particle interactions \cite{guo_control_2018}. REs in general form a vortex in phase space by the interactions of electric field acceleration, collisional damping, pitch-angle scattering, and synchrotron damping \cite{paz-soldan_resolving_2018}. The use of externally-launched waves to resonate with REs of particular energies provides a novel means of controlling this vortex in phase space. Pitch-angle scattering of REs to higher pitch-angle is desirable because the increased synchrotron damping they experience limits their energies, thus limiting their ability to penetrate into plasma-facing components, and potentially expose underlying structures, like cooling channels \cite{bartels_impact_1994}. It is the maximum energy of REs, not the maximum current, which best quantifies their great risk, and therefore limiting the maximum energy should be the aim of any effective mitigation strategy. Another benefit of limiting the energies of REs and not the current they carry is that sudden changes in current can induce large toroidal electric fields which can increase the avalanche growth rate. 
\par
An unbounded plane wave of frequency $\omega$ and wave vector $\mathbf{k}$ can resonate with an electron of positive classical gyro-frequency $\Omega$ travelling at velocity $\mathbf{v}$, relativistic factor $\gamma$, when the Doppler-shifted wave frequency matches an integer multiple of the cyclotron frequency:
\begin{equation}
    \omega - \mathbf{k} \cdot \mathbf{v} = l \Omega / \gamma,
    \label{eq:resonance}
\end{equation}
where the integers $l= -1, 0, +1$ correspond to the anomalous, Cherenkov (or Landau), and normal resonances, respectively. 

% Whenever the resonance condition, Eq. \ref{eq:resonance}, is met, a source of free energy is present, and the integral of the wave's electric field along a gyro-orbit of the electron is non-zero, energy can be transferred between the wave and the particle \cite{stix_waves_1992, koch_waveparticle_2006}.  
\par
Guo et al. \cite{guo_control_2018} have predicted that helicon waves can limit the energy of REs via the normal cyclotron resonance ($l=1$ in Eq. \ref{eq:resonance}). Helicon waves (also known as whistler waves, or fast waves in the lower hybrid range of frequencies) are right-handed electromagnetic waves that can propagate in plasmas but are evanescent in vacuum \cite{bhattacharjee_waves_2017, stix_waves_1992, pinsker_whistlers_2015}. The normal cyclotron resonance is the preferred resonance because it will interact with REs of an appropriate energy in the unmodified RE vortex. 

% These findings are consistent with the hypothesis that the waves are causing resonant pitch-angle scattering of high-energy REs, which are then damped down by synchrotron radiation to lower energies. Finally, RE growth rates across a wide range of shots with and without helicon are presented. This multi-shot database clearly shows that helicon waves dramatically reduce RE growth rates in plasmas that would otherwise experience exponential RE growth. 
% Finally, we compare two otherwise-identical shots which have different helicon directions and very different effects on the RE population. 
 \paragraph{Experimental Design}
The DIII-D tokamak is uniquely capable of investigating this novel RE mitigation strategy because it has both a high-power helicon antenna \cite{compernolle_high-power_2021, pinsker_experiments_2018, pinsker_first_2024}, and a low-density quiescent experimental scenario, capable of producing and controlling a benign RE population for several seconds \cite{spong_first_2018, heidbrink_low-frequency_2018, paz-soldan_spatiotemporal_2017, paz-soldan_resolving_2018}. It is also highly configurable: the directions of both the plasma current, $I_p$, and the toroidal magnetic field, $B_t$, can be changed independently.
\par
The DIII-D helicon antenna is a mega-watt level travelling-wave antenna that launches helicon waves in a \SI{10}{MHz} band centred at a frequency of \SI{476}{MHz}. The waves have a nominal parallel refractive index $n_\parallel = k_\parallel c / \omega = 3$ at the antenna \cite{compernolle_high-power_2021}, and can be launched in either toroidal direction. Strong antenna-plasma coupling has routinely been observed in DIII-D plasmas \cite{pinsker_first_2024} and strong coupling was experimentally verified for the low-density plasmas used for this experiment.
\par
In this experimental scenario, low-densities, $n_e \approx 0.5 \times 10^{19} \text{ m}^{-3}$, are used to achieve high normalized electric-fields, $E/E_C > 5$, enabling the growth of measurable and long-lived RE populations. These low-density plasmas require precise external error-field correction to prevent locked-modes from causing disruptions \cite{piron_error_2024, paz-soldan_non-disruptive_2022, paz-soldan_spectral_2014, scoville_locked_1991, callen_effects_2011}. 
\par
Two design decisions were important for this experiment: the direction of the toroidal magnetic field and the direction of the waves. During helicon operation, in addition to generating fast waves, there is some fraction of power launched directly by the antenna into the undesirable slow wave \cite{pinsker_whistlers_2015}. This fraction is minimized, however, when the direction of the total magnetic field is chosen to align with the Faraday screen of the antenna's modules. That alignment occurs when $B_t$ and $I_p$ are parallel. ($I_p$ is fixed to be anti-clockwise when viewed from above.) To satisfy the normal cyclotron resonance condition for REs with energies in the 5-10 MeV range with $\omega / (2\pi)=476$ MHz and $\Omega/ (2\pi)\approx6.5$ GHz, the wave and particle should approach each other, such that $\mathbf{k} \cdot \mathbf{v}<0$. To summarize, these design considerations lead to the experimental configuration in which the toroidal field, plasma current, and launched waves are all counter-clockwise when viewed from above; and REs travel clockwise when viewed from above.

% shown in Fig \ref{fig:experimentalsetup}. (As per convention, positive directions are parallel to the toroidal magnetic field.)

% \begin{figure}
%     \centering
%     \includegraphics[width=0.9\columnwidth]{images/optimal_experimental_setup.pdf}
%     \caption{Top down schematic view of the DIII-D tokamak. The radial axis shows the distance in metres from the tokamak centre. The pink region shows a plasma with a major radius of \SI{1.67}{m} and a minor radius of \SI{0.67}{m}. The direction of the bulk RE population, shown by $\mathbf{v_{RE}}$, is anti-parallel to the electric field which drives the plasma current, $I_p$. The direction of the helicon waves was chosen such that the main wave-particle resonance is the normal cyclotron resonance, made possible by $\mathbf{k} \cdot \mathbf{v_{RE}} < 0$. The direction of $B_t$ was chosen to minimise slow-wave excitation by the antenna. }
%     \label{fig:experimentalsetup}
% \end{figure}
\par
\begin{figure}
    \centering
    \includegraphics[width=0.9\columnwidth]{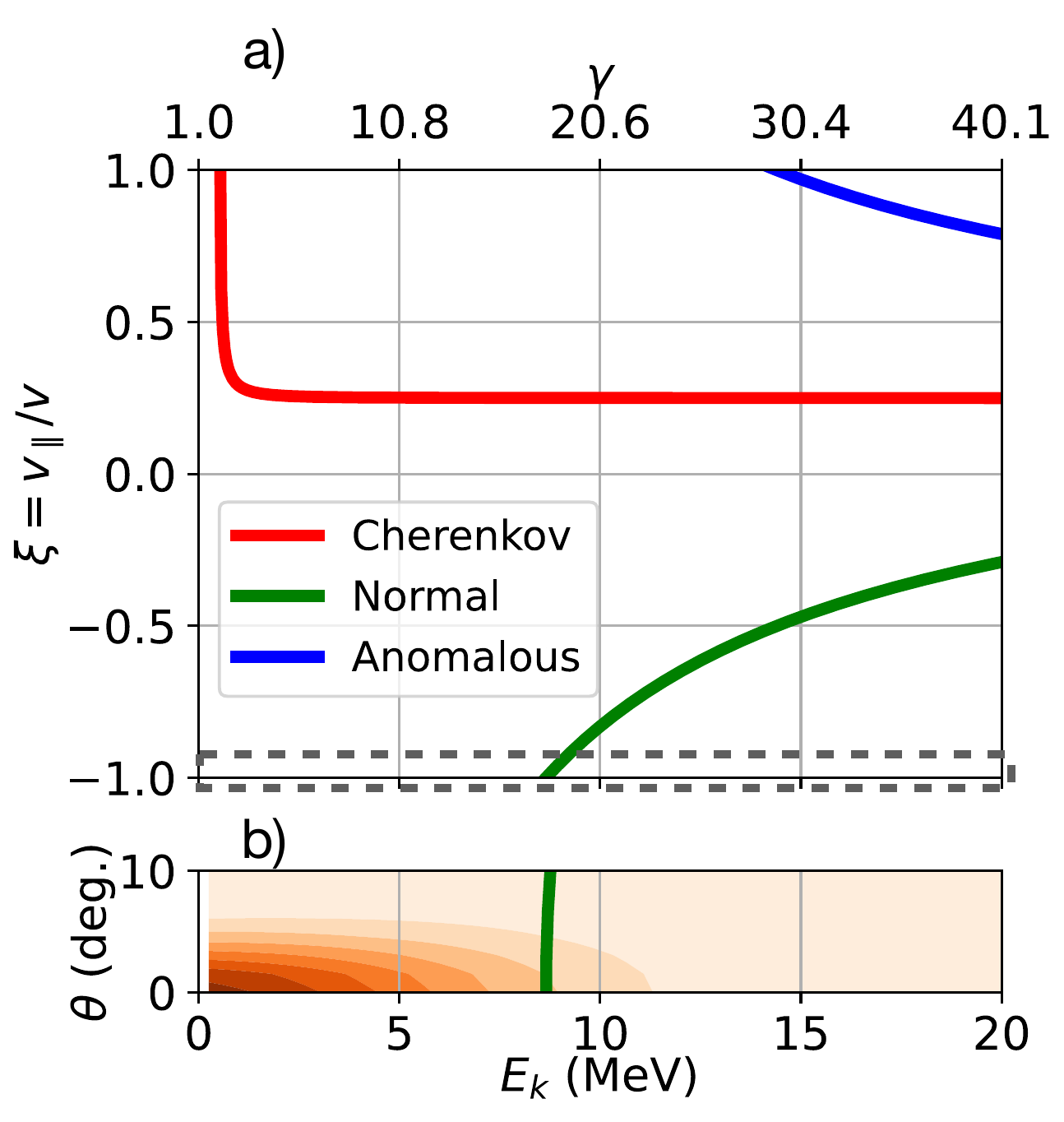}
    \vspace{-10pt}
    \caption{a) The resonances between helicon waves of 476 MHz and REs in phase space (pitch angle, $\xi = v_\parallel/v$, vs kinetic energy, $E_k$, in MeV, or equivalently the relativistic factor $\gamma$). Plasma parameters of shot 201948 have been used (i.e $n_e \approx 0.5 \times 10^{19} \text{ m}^{-3}, B_t=1.4$ T, and $k_\parallel = 40 \text{ m}^{-1}$). b) A representative normalized electron distribution function shown in the part of momentum space outlined by the dashed box in panel a). The pitch-angle shown in b) is the acute angle in degrees between the line of the magnetic field and the velocity of the REs.}
    \label{fig:resonance}
    \vspace{-15pt}
\end{figure}
Since the parallel velocity of a RE is much greater than its perpendicular velocity, the effect of $k_\perp$ in the resonance condition can be ignored (where $|\mathbf{k}|^2 = k_\perp^2 + k_\parallel^2$). To calculate $k_\parallel$, we note that owing to the toroidal symmetry of the tokamak, the toroidal mode number of the waves is conserved. This means as waves propagate inwards from the antenna on the low-field side to the tokamak interior, i.e. from larger to smaller major radius, the toroidal wavelength will decrease. This causes an up-shift in $k_\parallel$. Taking the antenna to be at a major radius of \SI{2.27}{m} and the magnetic axis to be at \SI{1.67}{m}, $k_\parallel$ increases by a factor of $2.27/1.67 \approx 1.36$. Since $n_\parallel = 3$ at the antenna and $c / \omega = (\SI{300}{\text{MHz m}}) / (2\pi \times \SI{476}{MHz}) \approx \SI{10}{m}$, this gives $k_\parallel = \SI{40}{\text{m}^{-1}}$ at the magnetic axis, and we use this in our calculations of the resonance condition ($\mathbf{k} \cdot \mathbf{v} = k_\parallel v_\parallel$). 
\par
Using these plasma parameters, the resonances in phase space are shown in Fig \ref{fig:resonance}. An inverse pitch-angle metric $\xi = v_\parallel / v = (\textbf{v} \cdot \mathbf{B_t}) / (|\mathbf{v}||\mathbf{B_t}|)$ is used (e.g., $\xi = 0$ for perpendicularly-traveling particles). REs in this experiment mostly travel antiparallel to $B_t$, i.e. close to $\xi = v_\parallel / v \approx -1$. This means low-pitch REs around 8 MeV resonate through the normal resonance. Panel b) in Fig \ref{fig:resonance} shows a representative normalized electron distribution function for this shot, in which $E/E_c =9$, $\hat{\tau}_{rad} = 10$, and $Z_{\text{eff}}=1.5$; the normalized synchrotron loss time, $\hat{\tau}_{rad}$, is calculated as $\left(6 \pi \varepsilon_0 m_0^3 c^3/ (e^4 B^2)\right)(1 / \tau)$, where $\tau \equiv 4 \pi \varepsilon_0^2 m_0^2 c^3 / e^4 n_e \Lambda$. Most REs travel within $6$ degrees of the background field with $E_k < $ \SI{10}{MeV}.
\par
The actual path of helicon waves is likely to be more complex than a simple single pass through the plasma. Once the waves are inside the plasma, the possibility of reflections and mode conversions cannot be ruled out. The plasma beta is too low to have any significant damping of the waves on the thermal plasma, so they are likely to make multiple passes through the plasma until they are absorbed by REs or mode converted. 
\par
% IGNORED PARA:
% In addition, the plasma parameters used are those of the plasma core and so the 0-D calculations of the resonance condition do not account for inhomogeneities or toroidal effects. To first order, particularly with the antenna's Faraday screen aligned with the magnetic field, and the vacuum gap between antenna and plasma minimized, most of the coupled helicon power is expected to travel as fast waves inside the plasma. %, parallel to the background field. 
% WAITING TO HEAR BACK FROM DON AND YASHIKA: A final point concerning the experimental design is that in a finite toroidal geometry, the resonances broaden and become more complex, so Eq. \ref{eq:resonance} and Fig \ref{fig:resonance} are first-order approximations of where the primary resonances will be in RE momentum space. 
\paragraph{Experimental Results}
Using the experimental configuration described earlier, helicon waves were applied across several shots and had a marked effect. To take a representative example, we show results for DIII-D shot 201948, in which helicon waves were launched into a plasma with REs. The helicon antenna was turned on at \SI{3.4}{s}, in a pulsed fashion, with \SI{100}{ms} on at a maximum coupled power of $\sim \SI{150}{kW}$ and \SI{100}{ms} off, repeated until the end of the shot. This waveform was chosen to allow multiple comparisons of conditions with and without helicon power. 
\par
REs emit Hard X-rays (HXR) through bremsstrahlung primarily in two ways: first, by colliding with background ions as they travel in the plasma, and secondly, by striking the first wall's plasma-facing components. 
The latter source is not significant in this experiment, as there are no sudden loss events. 
The HXR flux, measured by a plastic scintillator sensitive to MeV-range HXRs, therefore serves as an energy-weighted proxy for the confined RE population \cite{heidbrink_neutron_1986, heidbrink_low-frequency_2018}. 
\par
\begin{figure}
    \centering
    \includegraphics[width=\columnwidth]{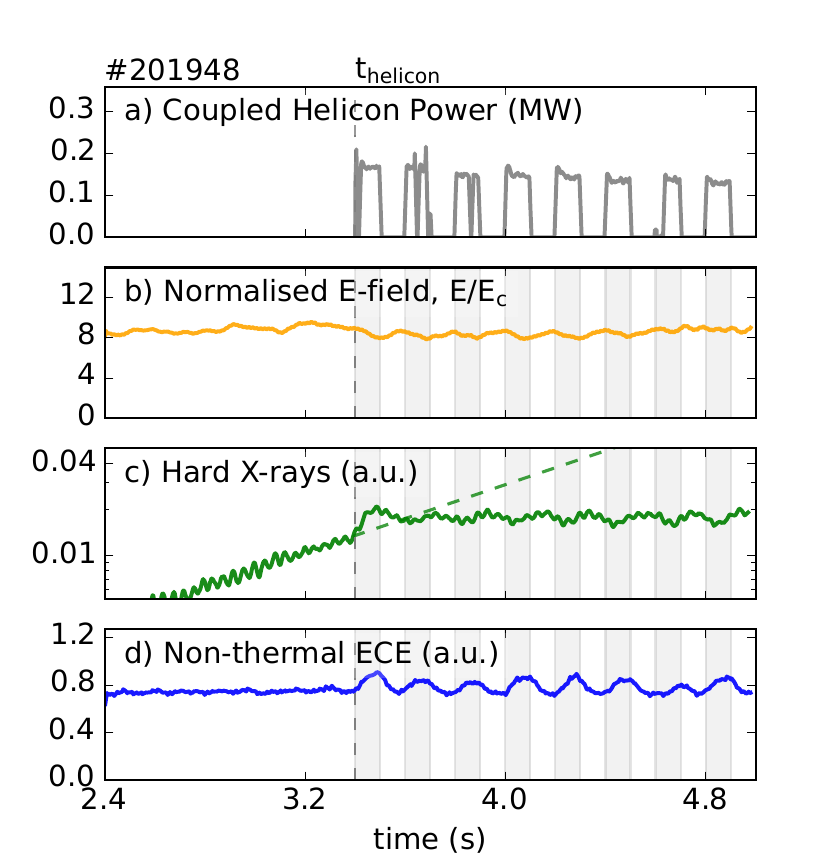}
    \vspace{-20 pt}
    \caption{Time traces of shot 201948. Panel (a) shows the application of 100 ms helicon pulses. The HXR signal, shown on a logarithmic scale in panel (c), flattens while the normalized electric field remains unchanged, panel (b). The dashed line in panel (c) shows an extrapolation of the HXR signal with the pre-helicon growth rate.  Panel (d) displays the non-thermal ECE signal, which increases by roughly 10\% to 20\% during helicon pulses.}
    \label{fig:timetraces}
    \vspace{-15pt}
\end{figure}
Fig \ref{fig:timetraces} shows key time traces for this shot.  The coupled helicon power is the power leaving the antenna (into the plasma), measured by an array of probes in the antenna. It does not account for the power-loss that takes place as the helicon waves traverse the vacuum region. For this shot, the average distance between the antenna and the last closed flux surface was \SI{4.5}{cm}. Prior to the application of helicon, the HXR-inferred RE population grows exponentially in time. Once helicon application starts, there are two timescales of interest: the effects during the \SI{100}{ms}-long helicon on-off pulses, and the longer behaviour over times greater than one second. On the shorter timescales during the \SI{100}{ms} helicon pulses, there are small increases to the HXR signal, with the largest increase happening during the first helicon pulse. The normalized electric field, $E/E_c$, also drops slightly. During the off periods, the HXR signal falls and $E/E_c$ increases slightly. These shorter timescale effects are very small. As will be shown later, the differences in $E/E_c$ cannot cause the pronounced change in RE-growth rate seen over the second or so following the start of helicon. Over many cycles (i.e., the next one to two seconds), the RE population flattens, if not decreases slightly. This flattening continues while the helicon power remains on. This is despite the normalized electric field being comparable to the pre-helicon phase. 
\par
REs also emit microwave electron-cyclotron emission (ECE) which is heavily Doppler-shifted and is detected on DIII-D using a 40-channel radiometer \cite{austin_electron_2003}. This non-thermal ECE signal is a strong function of the REs' perpendicular energies and is weighted towards low-energy REs \cite{liu_effects_2018, paz-soldan_resolving_2018}. The ECE signal, in Fig \ref{fig:timetraces} panel (d), increases when helicon is on, suggesting either an increase in low-energy REs or the average perpendicular energy, or a combination of the two.

\begin{figure}
    \centering
    \includegraphics[width=\columnwidth]{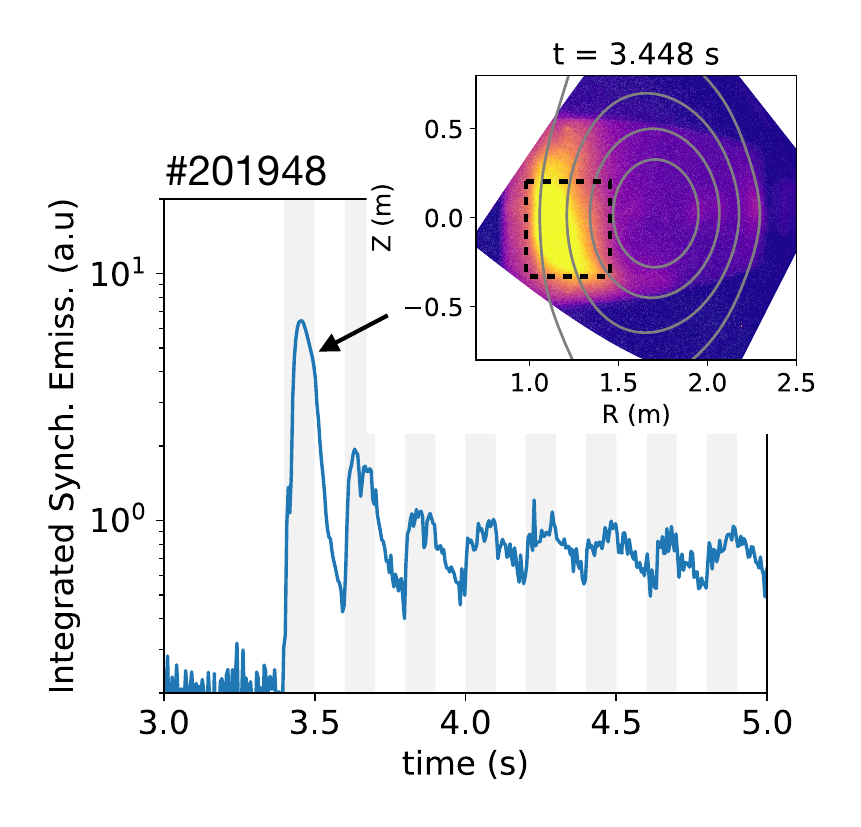}
    \vspace{-30 pt}
    \caption{Top inset, an R-Z image of a RE-synchrotron emission crescent, overlaid with magnetic flux contours. The bottom panel shows a time-trace of the fast camera signal integrated over the dashed black box in the inset. The grey shading shows when helicon power is on.}
    \label{fig:fastcam}
    \vspace{-15pt}
\end{figure}
\par
The synchrotron emission of REs, a strong function of the perpendicular velocity of the electrons \cite{hoppe_interpretation_2018}, is directly measured on DIII-D using a fast camera that captures images of light of wavelengths between \SI{700}{nm} and \SI{800}{nm}, at a rate of \SI{200}{Hz} \cite{marini_fast_2023}. The fast camera also shows a substantial change in the synchrotron emission when the helicon comes on. During the first \SI{100}{ms} of helicon pulse, a bright crescent in R-Z space appears on the high-field side, shown in the inset of Fig \ref{fig:fastcam}. These crescents are a standard feature of REs in tokamaks \cite{hoppe_interpretation_2018}. Integrating over the area on the high-field side outlined by the black-dashed line in the inset results in the time trace of the integrated synchrotron emission in the bottom panel of Fig \ref{fig:fastcam}. It is interesting to note that a crescent only appears during the first \SI{100}{ms} of helicon being on, and not subsequent pulses. One explanation of this is as follows: prior to the first pulse, there is an appreciable number of high energy REs ($>5$ MeV). When these are pitch-angle scattered by the helicon, they cause a very bright crescent. This population of high energy REs may not have time to recover, even during the \SI{100}{ms} between pulses. And so subsequent pulses of helicon fail to produce the bright crescent observed during the first pulse.
\par
\begin{figure}
    \centering
    \includegraphics[width=\columnwidth ]{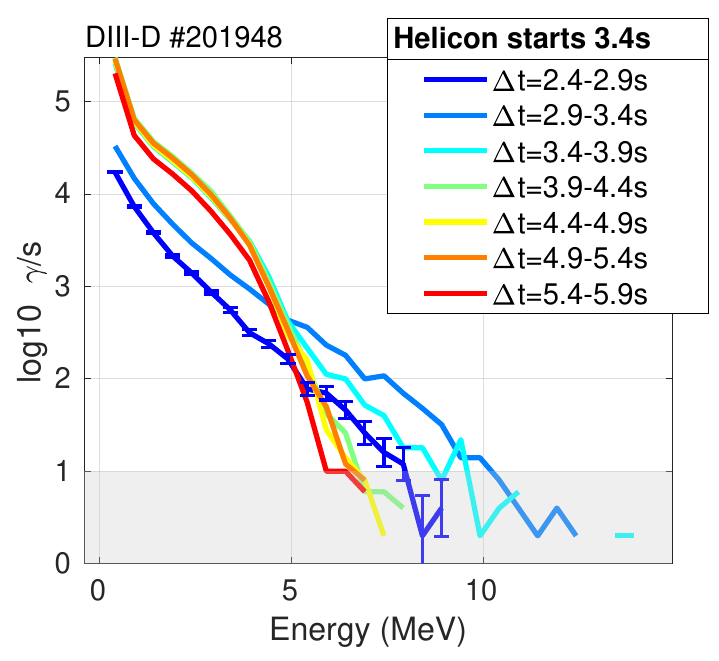} 
    \caption{The spectra of HXRs emitted by the REs for different time windows. The y-axis shows the normalized counts of photons in energy bins \SI{0.5}{MeV} wide. Representative errors from the counting statistics are shown for only the first \SI{500}{ms} spectrum for clarity \cite{wasserman_all_2004}. The grey shaded region at the bottom of the plot shows where the statistics for pulse-height analysis of the photons are less accurate.}
    \label{fig:gri}
    \vspace{-15pt}
\end{figure}

This explanation is supported by the time evolution of the energy distribution of HXRs, which is measured on DIII-D using an array of collimated detectors known as the Gamma Ray-Imager  \cite{pace_gamma_2016, cooperApplyingNewGamma2016, lvovskiyUpgradesGammaRay2022, paz-soldan_resolving_2018, paz-soldan_spatiotemporal_2017}. This uses pulse-height analysis over time windows of length $>$ \SI{200}{ms} to measure the energy spectrum of HXRs with energies greater than \SI{0.5}{MeV}. Energy spectra of HXRs integrated over time windows of \SI{500}{ms} each are shown in Fig \ref{fig:gri}. After helicon power starts at \SI{3.4}{s}, there is a significant reduction in high-energy ($>7$ MeV) REs and an increase in low-energy REs ($<4$ MeV). This supports the general hypothesis of resonant pitch-angle scattering, particularly since the reduction in high-energy REs agrees well with the predicted wave-particle resonance shown earlier in Fig \ref{fig:resonance}, and there is an increase in lower energy REs.
Coupled with the theoretical predictions, Figs \ref{fig:timetraces} and \ref{fig:fastcam} and \ref{fig:gri} strongly suggest that the maximum RE energy is being limited to less than $\sim 8$ MeV due to resonant pitch-angle scattering. 
% Specifically, the evidence suggests that a vortex in phase space has been formed in which REs are accelerated by the electric field until the wave-particle resonance at 8 MeV is satisfied, and then are pitch-angle scattered and then synchrotron damped down to lower energies, where the vortex starts again.

\begin{figure}
    \centering
    \includegraphics[width=0.9\columnwidth]{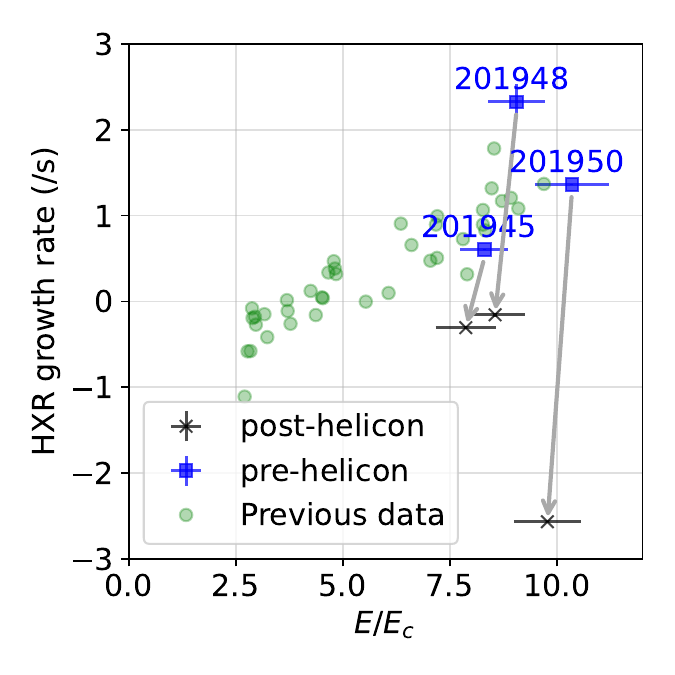}
    \vspace{-17 pt}
    \caption{The exponential HXR growth rate measured across a wide range of shots as a function of normalized electric field. Previous data shown are from Ref \cite{paz-soldan_growth_2014}. The post-helicon growth rates are calculated over the second following the start of helicon power. }
    \label{fig:database}
    \vspace{-15pt}
\end{figure}

\paragraph{Multi-shot database}
A multi-shot database has been compiled, relating the exponential growth rate of the RE population to the normalized E-field ($E/E_c$). It includes previous data from Ref \cite{paz-soldan_growth_2014}, which shows across dozens of shots that RE populations stopped growing or exponentially decayed roughly when $E/E_c \leq 5$ \cite{paz-soldan_growth_2014}. The growth rates versus normalized electric fields, including the new shots with helicon power, are shown in Fig \ref{fig:database}, and illustrate clearly that following the application of helicon waves, the RE population, as measured by the HXR signal, either ceases to grow or decreases exponentially. The post-helicon growth rates do not follow the same trend with normalized field as the other data. 

\paragraph{Conclusions}
% REs pose a formidable risk to future high-current tokamaks. It is therefore important that effective RE mitigation strategies exist which are able to limit the energy of the REs, since it is the energy that causes the most damage to plasma-facing components. A worst-case scenario that could shut a tokamak down for weeks involves 10-20 MeV REs directly striking a first-wall tile and exposing the underlying water cooling channel. 
% \par
This paper presents experimental evidence of the effectiveness of a novel RE mitigation strategy which involves launching helicon waves from an antenna in-vessel. This technique utilises the normal wave-particle resonance to pitch-angle scatter high energy REs which are then damped by an increase in synchrotron radiation back down to lower energies. 
\par
As illustrated in the case of DIII-D shot 201948 and seen across a number of shots,  the gamma-ray imager detects a drop in high-energy REs ($> \SI{7}{MeV}$) and an increase in low-energy REs ($< \SI{4}{MeV}$), just as the wave-particle resonance condition predicts. In addition, the hard x-ray signal plateaus, and sometimes decreases; the non-thermal electron-cyclotron emission increases with every pulse of helicon; the fast camera measures a brighter RE synchrotron emission crescent, most prominently for the first pulse of helicon. All these effects happen despite the normalized electric field, $E/E_c$, remaining high enough to generate REs without helicon. 
\par
Demonstrating this technique in reactor-relevant tokamak disruption scenarios remains a task for future work. Despite that, helicon waves appear worthy of consideration as a potential tool for RE mitigation, and their application in parallel with other strategies should certainly be considered moving forward. 

\paragraph{Acknowledgements}
The authors wish to thank J. Hanson for help on the plasma error-field correction, M.P. Ross for help programming the helicon waveforms, D. Liu for help setting up the hard x-ray scintillators, and R. Pinsker for helpful discussions about helicon wave propagation. 
\par
This material is based upon work supported by the U.S. Department of Energy, Office of Science, Office of Fusion Energy Sciences, using the DIII-D National Fusion Facility, a DOE Office of Science user facility, under Awards 
DE-FC02-04ER54698, % general DIIID and GA helicon
DE-SC0022270, % carlos DIIID
DE-SC0021622, % Frontiers
DE-SC0020337, % heidbrink 1
DE-AC05-00OR22725, % Don's
and  
DE-FG02-07ER54917. % claudio fast camera 
\par
We thank the U.S. Department of Energy Office of Fusion Energy Sciences through the Basic Plasma Science Program at Los Alamos National Laboratory (LANL) under contract No. 89233218CNA000001. This research used resources of the National Energy Research Scientific Computing Center, a DOE Office of Science User Facility supported by the Office of Science of the U.S. Department of Energy under Contract No. DE-AC02-05CH11231 using
NERSC award FES-ERCAP0032298 and LANL Institutional Computing Program, which is supported by the U.S. Department of Energy National Nuclear Security Administration under Contract No. 89233218CNA000001.
%{\fontsize{6pt}{3pt}\selectfont % Set font size to 6pt with a line spacing of 7.2pt
\par
Disclaimer: This report was prepared as an account of work sponsored by an agency of the United States Government. Neither the United States Government nor any agency thereof, nor any of their employees, makes any warranty, express or implied, or assumes any legal liability or responsibility for the accuracy, completeness, or usefulness of any information, apparatus, product, or process disclosed, or represents that its use would not infringe privately owned rights. Reference herein to any specific commercial product, process, or service by trade name, trademark, manufacturer, or otherwise does not necessarily constitute or imply its endorsement, recommendation, or favoring by the United States Government or any agency thereof. The views and opinions of authors expressed herein do not necessarily state or reflect those of the United States Government or any agency thereof.
\paragraph{Author Contributions}

H. Choudhury - Formal Analysis, Investigation, Methodology, Writing. 
A. Battey, C. Paz-Soldan - Supervision, Methodology, Investigation.
J. Lestz - Methodology, Investigation.
D. Spong, W. Heidbrink - Conceptualization. 
X. Tang, Q. Zhang, and Y. Zhang - Methodology.
The remaining authors all contributed equally to investigation and data curation. 

\bibliography{refs}

\end{document}